# Wave Propagation and Diffusive Transition of Oscillations in Pair Plasmas with Dust Impurities

Barbara Atamaniuk and Andrzej J. Turski

*Institute of Fundamental Technological Research, PAS, Świętokrzyska 21, 01-049, POLAND*

**Abstract.** In view of applications to electron-positron pair-plasmas and fullerene pair-ion-plasmas containing charged dust impurities a thorough discussion is given of three-component Plasmas. Space-time responses of multi-component linearized Vlasov plasmas on the basis of multiple integral equations are invoked. An initial-value problem for Vlasov-Poisson/Ampère equations is reduced to the one multiple integral equation and the solution is expressed in terms of forcing function and its space-time convolution with the resolvent kernel. The forcing function is responsible for the initial disturbance and the resolvent is responsible for the equilibrium velocity distributions of plasma species. By use of resolvent equations, time-reversibility, space-reflexivity and the other symmetries are revealed. The symmetries carry on physical properties of Vlasov pair plasmas, e.g., conservation laws. Properly choosing equilibrium distributions for dusty pair plasmas, we can reduce the resolvent equation to: (i) the undamped dispersive wave equations, (ii) wave-diffusive transport equation (iii) and diffusive transport equations of oscillations. In the last case we have to do with anomalous diffusion employing *fractional derivatives in time and space*. Fractional diffusion equations account for typical "anomalous" features, which are observed in many systems, e.g. in the case of dispersive transport in amorphous semiconductors, liquid crystals, polymers, proteins and biosystems.

## INTRODUCTION

The crucial point of the paper is the relation between equilibrium distributions of plasma species and the type of propagation (diffusive transition) of plasma response to a disturbance.

The paper contains a unified treatment of disturbance propagation (transport) in the linearized Vlasov electron-positron and fullerene pair plasmas, based on the space-time convolution integral equations.

We investigate the Vlasov-Ampère/Poisson system of equations for multicomponent plasmas, i.e.

(1) $$\left[\partial_t + u\partial_x + \frac{q_\alpha}{m_\alpha} E(x,t)\partial_u\right] F_\alpha(u,x,t) = 0, \quad \partial/\partial u = \partial_u \quad \text{(Vlasov)}$$

(2) $$\varepsilon_0 \partial_t E(x,t) + \sum_\alpha q_\alpha \int_{-\infty}^{\infty} u F_\alpha(u,x,t) du = 0, \quad \partial/\partial x = \partial_x, \quad \partial/\partial t = \partial_t \quad \text{(Ampère)}$$

(3) $$\varepsilon_0 \partial_x E(x,t) + \sum_\alpha q_\alpha \int_{-\infty}^{\infty} F_\alpha(u,x,t) du = 0, \quad E = -\partial_x \phi \quad \text{(Poisson)},$$

In view of (1), equations (2) and (3) are equivalent if appropriate constrains are applied to initial conditions for $F_\alpha$. Let us assume

(4) $$F_\alpha(u,x,t) \cong N_0^\alpha F_{0\alpha}(u) + F_{1\alpha}(u,x,t)$$

where $N_0^\alpha$, $F_{0\alpha}$ are the equilibrium particle concentration and the velocity distribution for $E = 0$, and $F_{1\alpha}$ is of the order $E$. Substituting (4) into (1), we derive the well-known linear equation:

(5) $\quad (\partial_t + u\partial_x)F_{1\alpha} = -(N_0^\alpha q_\alpha / m_\alpha)E\,\partial_u F_{0\alpha}$

For the initial-value problem

(6) $\quad F_{1\alpha}(u,x,0) = g_\alpha(u,x)$, $\quad g_\alpha(u, x = \pm\infty) = 0 \quad and \quad E(x,t) = 0 \quad for \quad t \leq 0$,

we have the solution of (5) and using (2), we write

(7) $\quad E(x,t) = \int_0^t dt_1 \int_{-\infty}^\infty K(\xi, t_1) E(x-\xi, t-t_1) d\xi + G(x,t)$

where

$$G(x,t) = -\sum_\alpha (q_\alpha / m_\alpha) \int_0^t \int_{-\infty}^\infty u g_\alpha(u, x-ut_1) du\, dt_1, \quad K(x,t) = -\sum_\alpha \omega_\alpha^2 F_{0\alpha}(x/t) \quad and \quad \omega_\alpha^2 = \frac{N_0^\alpha q_\alpha^2}{\varepsilon_0 m_\alpha}.$$

More detailed derivation of (7) can be found in [1 and 2].

## RESOLVENT KERNEL EQUATIONS

The space-time convolution equation (7) can be solved by use of a resolvent kernel $R(x,t)$. We shall write the solution as

(8) $\quad E(x,t) = G(x,t) + \int_0^t dt_1 \int_{-\infty}^\infty G(\xi, t_1) R(x-\xi, t-t_1) d\xi$.

Where the forcing functions $G(x,t)$ and the resolvent $R(x,t)$ satisfies the following resolvent equation

(9) $\quad R(x,t) = K(x,t) + \int_0^t dt_1 \int_{-\infty}^\infty K(\xi, t_1) R(x-\xi, t-t_1) d\xi$.

The last equation describes the plasma dynamic response $R(x,t)$ and its only dependence on the plasma equilibrium distribution $K(x,t)$. It is evident that for the infinite support $x \in (-\infty, \infty)$ of a kernel $K(x,t)$, the $R(x,t)$ also possesses the infinite support. The physical consequence of the property is that the plasma response to any disturbance, even if the disturbance is with a limited support, appears in the full space $x \in (-\infty, \infty)$. On the ground of (9) we note, that for $K(x,t) = K(x,-t)$ it follows that $R(x,t) = R(x,-t)$ and for $K(x,t) = K(-x,t)$ we have $R(x,t) = R(-x,t)$. The property is reversible with respect to $R(x,t)$ and $K(x,t)$. It is called the time reversibility and space reflexivity. The important point to note here is that according to the Noether theorem the properties are strictly related to energy and momentum conservation laws.

The time-Laplace and space-Fourier transforms of (9) lead to the usual dispersion relation of multicomponent plasmas- $D(k,s)$

(10) $\quad R(k,s) = \dfrac{K(k,s)}{1 - K(k,s)} \quad and \quad D(k,s) \equiv 1 - K(k,s) = 0$, where $D(k,s)$ is the Fourier-Laplace symbol. In the case of diffusive transport equation of oscillations the relation has no meaning. It is worth pointing out that the resolvent equation is more universal description of multicomponent plasmas than the usual dispersion relations.

## WAVE PROPAGARTIONS

The advantage of the integral equations of Vlasov plasmas consists in obtaining the solutions separately composed of the forcing function $G(x,t)$ resulting from the initial value disturbance $g(u,x)$ and the resolvent kernel depending only on the plasma equilibrium $\sum_{\alpha} F_{0\alpha}(u)$. Assuming the hot components of *pair plasma* with the so-called *"square"* equilibrium velocity distributions and the cold heavy dust grains, we have:
$K(x,t) = -\omega_d^2 t \delta(x) - (\omega_g^2/2a)[H(x+at) - H(x-at)]$, where $\omega_d^2 = N_d q^2/\varepsilon_0 m_d$ and the effective gap frequency is: $\omega_g^2 = (N_0 q^2/\varepsilon_0 m_0)(2-\nu)$. The constant $\nu$ is to ensure the charge neutrality of the plasma.

Hence the Fourier and Fourier-Laplace symbols are
$$K(k;t) = -\omega_d^2 t - (\omega_g^2/ak)\sin(kat) \quad \text{and} \quad K(k;s) = -\frac{\omega_d^2}{s^2} - \frac{\omega_0^2}{s^2 + k^2 a^2}.$$

The dispersion relation takes the form
$$D(k;s) \equiv s^4 + s^2 k^2 a^2 + s^2(\omega_d^2 + \omega_h^2) + \omega_d^2 k^2 a^2 = 0.$$

The respective dust-pair plasma wave equation for the resolvent kernel takes the form:
(11) $\qquad R_{tttt} - a^2 R_{xxtt} + (\omega_d^2 + \omega_g^2) R_{tt} - \omega_d^2 a^2 R_{xx} = 0$

and we note, that $R(x,t)$ are time reversible and x-space reflective.

Assuming that $\omega_d^2/\omega_g^2 \ll 1$ and introducing the dust acoustic velocity $c_s = a\omega_d/\omega_g$, we can approximate the last equation, as follows;
(12) $\qquad R_{tt} - c_s^2 R_{xx} - (a^2/\omega_g^2) R_{xxtt} = 0$.

Hence, we can expect dust acoustic waves in the pair plasma with dust grains. Also, one can expect dust acoustic solitons in the case of finite amplitude and nonlinear waves.

## MAXWELLIAN PLASMAS

Maxwellian equilibrium distribution
(13) $\quad F_{0\alpha}(u) = a_\alpha \pi^{-1/2} \exp(-a_\alpha^2 u^2), \quad \alpha = e, p, \quad <u_\alpha^2> = 1/2a_\alpha^2$

is considered to be most appropriate but analytically almost intractable and the exact closed-form solution of equation (9) is not yet known, see [5]. In the paper [2], Maxwellian plasmas are analyzed by means of approximate formulae and computer diagram presentations. The main results of the computation concerning the Maxwellian plasmas can be summarized as follows. The nature of plasma response is a compound of a diffusive transition, being essentially a plasma oscillation mode with the $\omega_0$ – plasma frequency and the Gaussian types of amplitude envelop, and a decreasing dispersive wave mode. Differentiation of these two properties is not an easy task and we have not a ready conclusion but it seems that the Maxwellian plasma response exhibits mainly diffusive transport in space for fixed values of time in a long time range, and damped wave behavior with respect to time for fixed values of $x$. Such behavior is typical for the anomalous diffusion described by fractional derivative operators, see [4]. The presence of dust impurities requires new additional numerical calculations.

## ANOMALOUS DIFFUSION OF OSCILLATIONS

The next exact solution known to us is the resolvent for the Lorentz (Cauchy) pair plasma with dust impurities.. The equilibrium distribution is
$$F_{0+}(u) = F_{0-}(u) = \frac{\lambda}{\pi(\lambda^2 + u^2)} \quad \text{where } \lambda \text{ is a positive number.}$$
The distribution is related to Lévy stable nongaussion processes and has no higher moments, e.g. mean-square velocity. It can be related to *anomalous diffusion* processes and is useful for modeling plasma with a high-energy tails that are typical in space plasmas.

We quote new results concerning the resolvent for pair plasma with dust grains.
Let us describe the kernels due to equilibrium distributions of plasma species:

(14)  $K(k;t) = -\omega_d^2 t - \omega_g^2 t \exp(-|k|\lambda t)$,  $K(x,t) = -\omega_d^2 t \delta(x) - (\omega_g^2/\pi)\lambda/(\lambda^2 + x^2/t^2)$

The Laplace-Fourier symbol of the resolvent kernel takes the form:

(15)  $$R(k,s) = -\frac{\omega_d^2(s+|k|\lambda)^2 + s^2\omega_g^2}{(s^2+\omega_d^2)(s+|k|\lambda)^2 + s^2\omega_g^2}$$

Introducing the parameter $\epsilon = \omega_d^2/\omega_g^2 < 1$, we can write

(16)  $R(k,s) = R_0(k,s) + \sum_{n=1}^{\infty} \epsilon^n R^c{}_n(k,s)$ and $R(x,t) = R_0(x,t) + R_d(x,t) + \sum_{n=1}^{\infty} \epsilon^n R_n(x,t)$.

Where $R_0(k,s) = -\dfrac{\omega_g^2}{\omega_g^2 + (s+|k|\lambda)^2}$, hence $R_0(x,t) = -\dfrac{\lambda}{\pi}\dfrac{\omega_g t \sin \omega_g t}{(\lambda t)^2 + x^2} = -\omega_g \rho(x,t)\sin\omega_g t$.

The oscillating component with the dust plasma frequency is: $R_d(x,t) = -\omega_d \delta(x)\sin\omega_d t$.

The higher order terms due to the dust presence:

$$R_n(k,s) = (-1)^n \frac{\omega_g^{2n}}{s^{2n}}\left\{[1 - \frac{\omega_g^2}{[\omega_g^2 + (s+|k|\lambda)^2]}]^{n+1} - 1\right\}$$

hence $R_n(x,t) = (-1)^n \int_0^t g_n(t-t_1)\rho(x,t_1)\Phi_n(t_1)dt_1$,

where $g_n(t) = \omega_g^{2n} t^{2n-1}$, $\rho(x,t) = \dfrac{1}{\pi}\dfrac{\lambda t}{(\lambda t)^2 + x^2}$

and $\Phi_n(t) = L^{-1}\{(1-\dfrac{\omega_g^2}{\omega_g^2+s^2})^{n+1} - 1\} = P_n(\omega_g t)\sin\omega_g t + Q_n(\omega_g t)\cos\omega_g t$. The $P_n(.)$ and $Q_n(.)$ are polynomials with respect to $\omega_g t$. For example, the second term is as follow:

$\Phi_2(t) = \omega_g(-\dfrac{3}{2}\sin\omega_g t - \omega_g t \cos\omega_g t)$

The resolvent $R(x,t)$ is drastically different from the previous one. It does not exhibit wave propagation and there is no dispersion relation. We observe a *"diffusive transition"* of oscillations. The amplitudes (envelops) $\rho(x,t)$ of a bit more complicated oscillations does obey more complex fractional diffusive equation. Wave damping has no meaning, but time reversibility and space reflexivity are preserved.

If we assume the solution in the form

(17)  $R(x,t) = -\omega_0 \rho(x,t)\sin(\omega_0 t)$, where $\int_{-\infty}^{\infty}\rho(x,t)dx = 1$,

then the resolvent equation (9) takes the form

(18)  $F_0(x,t) - t\rho(x,t) = \omega_0 \int_0^t dt_1(t-t_1)\sin(\omega_0 t_1) * \left[\int_{-\infty}^{\infty}\rho(x_1,t_1)\dfrac{F_0(x-x_1,t-t_1)}{t-t_1}dx_1 - \rho(x.t)\right]$

We denote $K(x,t) = -\omega_0^2 F_0(x,t)$ and use the relation

$\sin(\omega_0 t) = \omega_0 t - \omega_0^2 \int_0^t (t-t_1)\sin(\omega_0 t_1)dt_1$.

For more details we refer the reader to [1 and 2].

If $\rho(x,t) = (1/t)F_0(x,t)$, then the resolvent equation implies the following Chapman-Kolmogoroff equation

(19) $\rho(x,t) - \int \rho(x-x_1, t-t_1)\rho(x_1,t_1)dx_1 = 0$

The equation has a unique solution if and only if the following integral exists:

$\int_{-\infty}^{\infty} x^2 \rho(x,t)dx = 2Dt$ and the solution is the Gaussian (normal) probability density distribution. In the case that the integral does not exist, then the equation (18) can posses many different solutions. Let the function $\rho(x,t)$ satisfies the following Cauchy relation (1850)

$\int_{-\infty}^{\infty} e^{ikx}\rho(x,t)dx = \exp(-|k|^\alpha t)$ , where $0 < \alpha < 2$, and $\exp(-|k|^\alpha t)$ is a characteristic function (Fourier transform), hence $\rho(x,t)$ is the Lévy $\alpha$-stable probability distribution function. The function satisfies the Chapman-Kolgomoroff equation (19). Since $\rho(k;t) = \exp(-|k|^\alpha t)$, we see at once that the characteristic function is a solution to the following equation

$\dfrac{d}{dt}\rho(x,t) = -|k|^\alpha \rho(k;t)$  with $\rho(k;0) = 1$. Inverting shows that the distribution function $\rho(x,t)$ solves a fractional partial differential equation, (fractional diffusion equation)

$$\partial_t \rho(x,t) = \partial_{|x|}^\alpha \rho(x,t)$$

The symmetric fractional derivative operator $\partial_{|x|}^\alpha$ corresponds to multiplication by the symbol -$|k|^\alpha$ in the Fourier space. For more details on symmetric $\alpha$-stable ($S\alpha S$) processes we refer the reader to [6 and 7].

## CONCLUSIONS

- An initial-value problem for Vlasov-Poisson/Ampère equations has been reduced to the integral equation and the solution to the problem is expressed in terms of a forcing function $G(x,t)$ and its convolution with a resolvent kernel $R(x,t)$.
- The forcing function is responsible for the initial disturbance and the resolvent is responsible for equilibrium distributions. Resolvent kernel equations are eligible for computer calculations.
- We have exhibited three types of exact closed-form solutions of the space-time resolvent equations. These solutions can be classified following the space-time behavior.
- Dust impurities may cause appearance of dust acoustic waves and solitons. They disturb oscillations but the diffusive transitions remain unchanged according to envelop $\rho(x,t)$.
- There is a suggestion that the envelopes of diffusive transition of oscillations can be governed by a symmetric $\alpha$-stable (S$\alpha$S) process. The probability distributions of the processes are related to the fractional diffusive transition described by the fractional diffusion equations.

**This research is supported by KBN grant nr 0TOOA01429**

## References

1. A. J. Turski, *Il Nuovo Cimento*, Serie X, Vol. **63B**, pp. 115-131, (1969).
2. A. J. Turski and J. Wojcik, *Arch. Mech.* **48**, 1047-1067, Warsaw (1996).
3. A. J. Turski, B. Atamaniuk and E. Turski "ArXiv: Math-ph/0701068.
4. A. J. Turski and B. Atamaniuk, "Wave Propagation and Diffusive Transition of Oscillations in Pair Plasmas" in *Plasma 2007, Greifswald (Germany),16-19 October 2007*, edited by H. J. Hartfuss et. al., AIP Conference Proceedings 993, May 2008, pp. 109-112.
5. A. J. Turski, An integral equation of convolution type, SIAM Review, **10,** 1, 108, 1968.

6. A. J Turski, B. Atamaniuk and E. Turska, *Application of Fractional Derivative Operators to Anomalous Diffusion and Propagation Problems,* ArXiv: Math-ph/0701068
7. G. Samorodnitski and M. S. Taqqu, *Stable non-Gaussian random processes* , Chapman & Hall, N. York-London, 1994